\documentclass[pra,10pt,twocolumn,superscriptaddress,floatfix,showpacs]{revtex4-1}

\usepackage[utf8]{inputenc}  
\usepackage[T1]{fontenc}     
\usepackage[british]{babel}  
\usepackage[sc,osf]{mathpazo}
\usepackage[scaled=0.86]{berasans}  
\usepackage[colorlinks=true, citecolor=blue, urlcolor=blue]{hyperref}  
\usepackage{graphicx} 
\usepackage[babel]{microtype}  
\usepackage{amsmath,amssymb,amsthm,bm,amsfonts,mathrsfs,bbm} 

\usepackage{xspace}  
\usepackage{pgfplots}

\begin{document}

\title{Study of nonlocal correlations in macroscopic measurement scenario}


\author{Samir Kunkri}
\affiliation{Mahadevananda Mahavidyalaya, Monirampore, Barrakpore, North 24 Parganas-700120, West Bengal, India.}

\author{Manik Banik}
\affiliation{Optics \& Quantum Information Group, The Institute of Mathematical Sciences, HBNI, C.I.T Campus, Tharamani, Chennai 600 113, India.}

\author{Sibasish Ghosh}
\affiliation{Optics \& Quantum Information Group, The Institute of Mathematical Sciences, HBNI, C.I.T Campus, Tharamani, Chennai 600 113, India.}


\begin{abstract}
Nonlocality is one of the main characteristic features of quantum systems involving more than one spatially separated subsystems. It is manifested theoretically as well as experimentally through violation of some \emph{local realistic} inequality. On the other hand, classical behavior of all physical phenomena in the macroscopic limit gives a general intuition that any \emph{physical} theory for describing microscopic phenomena should resemble classical physics in the  macroscopic regime-- the so-called macro-realism. In the 2-2-2 scenario (two parties, each performing two measurements, each measurement with two outcomes), contemplating all the no-signaling correlations, 
we characterize which of them would exhibit \emph{classical} (\emph{local-realistic}) behaviour in the macroscopic limit. Interestingly, we find correlations which at single copy level violate the Bell-Clauser–Horne-Shimony–Holt inequality by an amount less than optimal quantum violation (i.e., Cirel'son bound $2\sqrt{2}$), but in the macroscopic limit gives rise to a value which is higher than $2\sqrt{2}$. Such correlations are therefore not considered as physical.
Our study thus provides a sufficient criterion to identify some of unphysical correlations.  
\end{abstract}


\maketitle

\section{Introduction}
In our everyday experience almost all the physical phenomena satisfy the laws of the classical physics. However, at the microscopic scale the physical world follows the rules of quantum physics. The description of quantum physics is different from its classical counterpart both conceptually as well as mathematically \cite{Ballentine}. This raises the question of quantum to classical transition, i.e., when and how do the systems stop behaving quantum mechanically and begin to behave classically? Several novel ideas, like collapse models \cite{Collapse}, concept of decoherence \cite{Zurek} \emph{etc} have been introduced long back to address these questions. More recently, in a conceptually different approach, it has been shown that under coarse-grained measurements, classical world arises out of quantum physics \cite{Kofler'2007}. All these studies result into a general dictum that in the macroscopic level, the non-classical behaviors of quantum theory or any physical theory (possibly post quantum) should subside, and consequently classicality should emerge. Aim of this paper is to study the emergence of such classical behavior in terms of strength of correlations for generalized no-signaling theories and identify some of the generalized no-signaling correlations as \emph{unphysical}.   

One of the most fundamental contradictions of quantum mechanics (QM) with classical physics is its nonlocal behavior as established by J. S. Bell in his 1964 seminal work \cite{Bell'64} (see also \cite{Brunner'14}). Whereas all correlations in the classical world are \emph{local-realistic}, correlations obtained from multipartite entangled quantum systems may violate empirically testable local realistic inequality (called `Bell type inequalities' in general) which establishes that such quantum correlations do not allow a \emph{local realistic} explanation. Quantum nonlocality does not contradict the relativistic causality principle, or more generally the no-signaling principle. Moreover, QM is not the only possible theory that exhibits nonlocality along with satisfying the no-signaling principle; there can be non-quantum no-signaling correlations exhibiting nonlocality. One extreme example of such a correlation (more nonlocal than QM) was first constructed by Popescu and Rohrlich (PR) \cite{Popescu'94}. Whereas the PR correlation violates the Bell–Clauser–Horne-Shimony–Holt (Bell-CHSH) \cite{Clauser'69} inequality by algebraic maximum, the optimal Bell-CHSH violation in quantum theory is restricted by the Tsirelson’s bound \cite{Tsirelson'80}. This raises another important questions: which nonlocal correlations are physical? This question is also important from practical perspective since nonlocality has been proved to be important resource in numerous applications \cite{Applications1,Applications2,Applications3, Applications4,Applications5,Applications6, Applications7,Applications8,Applications9,Applications10}. An endeavor to answer this question was initiated by W. van Dam who showed that existence of super-strong nonlocal correlations [eg. PR correlation] would trivialize the problem of communication complexity \cite{vanDam'05}. It may be noted that principles like information causality (IC) \cite{Pawlowski'09} or macroscopic locality (ML) \cite{Navascues'09} do help us towards understanding physicality of some of the post-quantum correlations. Apart from these other conceptually different proposals have been introduced to single out the Tsirelson’s bound \cite{Others1,Others2,Others3,Others4,Others5}. But, till date, identifying the boundary between quantum correlations and post-quantum ones is not done completely and it remains an active area of research (see \cite{Zhen'16}). Here we aim at approaching this problem using a macroscopic measurement scheme different from the one used in ML. 

In order to study macroscopic properties of a correlation, one must take a measurement scheme on many copies of the correlation where the identity of individual particles involved in the correlation should not be revealed \cite{kurzynski'16}. A practically relevant scheme for studying such macroscopicity of correlations is to consider a case when the identity of the individual particles in the correlations gets lost during the distribution of the correlated state. One can of course interact microscopically with particles in the correlation, but in general, it is difficult to address them individually \cite{self0}. So whatever microscopic interaction one intend to use, it will affect, in general, all the particles of the beam at the same time. In this context, Bancal {\it et al.} have studied the violation of Bell inequalities of entangled states considering a general multipair scenario \cite{Bancal'08}. They have shown that the nonlocality of the quantum entangled state decreases in this multipair scenario with the increase in the number of independent entangled pairs, i.e., in the macroscopic limit of having infinitely many copies of entangled pairs, one can not get nonlocal correlation. This observation is compatible with the general dictum that classicality emerges in the macroscopic level. 

Here, in the simplest scenario, i.e., two parties, each performing one of the two possible measurements, and each measurement having the two possible outcomes (i.e. $2-2-2$ scenario), we consider the same approach as that of Bancal {\it et al.} \cite{Bancal'08}. But instead of considering only correlations in entangled quantum states, we contemplate general correlations that may be stronger than quantum ones in exhibiting nonlocal behavior, yet weak enough to prohibit instantaneous signaling. We characterize {\it all} such correlations, which, in the macroscopic limit, display classicality that is considered- in our context- to be the \emph{local realistic} behavior of the correlations. It is worth mentioning that such classical behavior of any correlation at macroscopic is not sufficient to certify the correlation to be perceived in some physical theory, it is rather a necessary criterion. We find examples of such correlations that in the macroscopic level behave classically but do not fulfill other necessary criteria, like \emph{nonlocality distillation} \cite{Forster'09,Allcock'09,Brunner'09,Hoyer'10} or IC \cite{Pawlowski'09}, and hence cease to be considered as physical correlations. Interestingly, on the other hand, we find examples of correlations that indeed satisfy the necessary criteria of IC but at the macroscopic scale exhibit strong nonlocal behavior, going against our general dictum, and hence fail to be considered as physical correlations.

At this point it is important to note that ML principle \cite {Navascues'09} can also identify unphysical correlations. However our approach is different from that of ML. In ML, one also consider beams of correlated particles ($M$ in number  which is much greater than $1$, i. e. in the thermodynamic limit) and to make a relationship with classical physics these beams are assumed to be continuous fields. In other words, the detectors in Alice's and Bob's can only perform coarse-graining measurements, i.e., these detectors cannot resolve the beams to their constituent particles (also, this implies that, they cannot perform different measurements on different particles, they are eligible to perform same interactions on the whole beam). Hence, the resolution of their detectors should not be perfect, and such a poor resolution can only provide the information about the mean value. The detectors work in such a precision that one could observe the deviations of intensity fluctuations from the mean value of the order $\sqrt{M}$, because in that case the resultant distributions will be described by classical physics. 

On the other hand, very recently Rohrlich showed that, at the macroscopic scale, PR-box correlations violate relativistic causality and hence has no realization in the classical world \cite{Rohrlich'13}. Moreover, this result has been generalized to all stronger-than-quantum bipartite correlations, constituting a derivation of Tsirelson’s bound without assuming quantum mechanics \cite{Rohrlich'13,Gisin'14}. While in Refs. \cite{Rohrlich'13,Gisin'14} the authors showed unphysicality of stronger than quantum correlation via showing signaling of those correlations in the macroscopic limit, we do the same, but, via distillation of non-locality in the macroscopic limit, and therefore our approach is completely different from that adopted in \cite{Rohrlich'13,Gisin'14}. Furthermore in our approach some of the  correlations having weaker nonlocality than the optimal quantum nonlocality (in the sense that Bell-CHSH violation is strictly less than Tsirelson’s bound) turns out to be unphysical.

Organization of our paper goes as follows: in section-(\ref{setup}) we discuss the setup to study a general bipartite correlation in single-pair as well as in multipair scenario; in section-(\ref{result}) we present our results and section-(\ref{comparison}) contains comparative discussion between our procedure and other methods; lastly we present our conclusion in section-(\ref{conclusion}).   
\begin{figure}[t!]
	\centering
	\includegraphics[height=3.3cm,width=7.5cm]{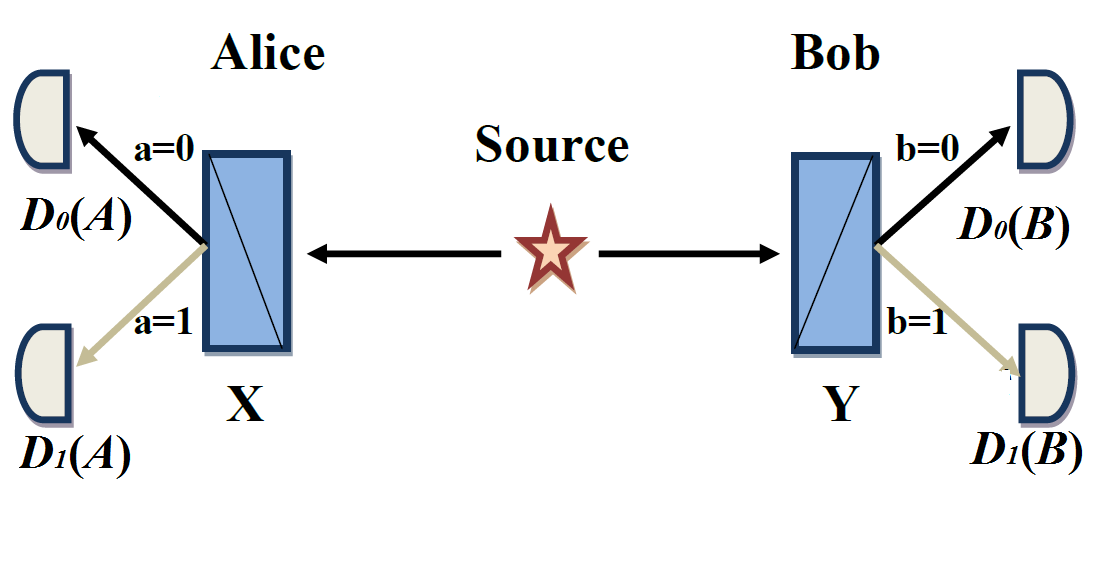}
	\caption{(Color on-line) Single-pair setup: $X,~Y\in\{0,1\}$ are Alice's and Bob's measurements, respectively. After the measurement interaction, going through the path $a=0$ and $a=1$, particles are collected at Alice's detectors $D_0(A)$ and $D_1(A)$, respectively, and similarly at Bob's end at the detectors $D_0(B)$ and $D_1(B)$.}\label{fig0}
	\vspace{-.5cm}
\end{figure}

\section{Setting up the scenario}\label{setup}
\subsection{Single-pair setting }
Consider the following bipartite scenario: a particle pair is produced by some source and two spatially separated experimentalists (say Alice and Bob) receive one particle each. Alice (Bob) can have one of the following two interactions denoted by $X = 0, 1$ $(Y = 0, 1)$ with her (his) particle. Each interaction results in Alice's (Bob's) particle to follow one of the two possible paths, called outcomes and let us denote it by $a$ ($b$) with $a\in\{0,1\}$ ($b\in\{0,1\}$), and eventually will impinge on one of Alice's (Bob's) two detectors $D_a(A)$ ($D_b(B)$ ) [see Fig.\ref{fig0}]. Repeating this experiment many times they can estimate the relative frequencies $P(ab|XY )$, i.e., the probability that Alice's and Bob's outcomes are $a$ and $b$, respectively when they apply the interactions $X, Y$. The joint probabilities $\{P(ab|XY)\}$ form an entire correlation vector. The positivity, normalization and non-signaling constraints lead to this correlation vector to be a point of an eight dimensional polytope \cite{Barrett'05}, called no-signaling polytope $\mathcal{NS}$. Local correlations are of the form $P(ab|XY)=\int d\lambda\rho(\lambda)P(a|X,\lambda)P(b|Y,\lambda)$, where $P(a|X,\lambda)$ is the probability of getting the outcome $a$ when Alice performs the measurement $X$ given the knowledge of (local hidden) variable $\lambda$, $P(b|Y,\lambda)$ is similar for Bob and  $\rho(\lambda)$ is a probability distribution over the variable $\lambda$. Collection of all such local correlations form another polytope $\mathcal{L}$ strictly residing in the $\mathcal{NS}$ with trivial facets determined by positivity constraints and nontrivial facets determined by Bell-CHSH inequalities, that up-to 
relabeling of inputs and outputs, read:
\begin{equation}\label{eq1}
I_{CHSH}:= |\langle 00\rangle+ \langle 01\rangle + \langle 10\rangle -\langle 11\rangle|\leq 2,
\end{equation}
where $\langle XY\rangle:=\sum_{a,b}(-1)^{a\oplus b}P(ab|XY)$, and $\oplus$ denotes modulo-2 sum. Correlations that are of the form $P(ab|XY)=\mbox{Tr}[\rho_{AB}(\Pi^a_X\otimes\Pi^b_Y)]$ are called quantum, where $\rho_{AB}$ is some density operator on some composite Hilbert space and $\{\Pi^a_X\}$ ($\{\Pi^a_X\}$) is some positive operator valued measure on Alice's (Bob's) side. The set of quantum correlations, $\mathcal{Q}$ forms a convex set (with continuous boundary) lying strictly between $\mathcal{NS}$ and $\mathcal{L}$, i.e., $\mathcal{L}\subset\mathcal{Q}\subset\mathcal{NS}$. There are $24$ vertices of the polytope $\mathcal{NS}$, $16$ of which are the extreme points of the polytope $\mathcal{L}$, called local/deterministic vertices and remaining $8$ are called nonlocal vertices. Since $\sum_{a,b}P(ab|XY)=1$ (due to normalization) hence $I_{CHSH}$ can be written as,
\begin{equation}\label{eq2}
I_{CHSH}= |2+2(A_{11} - A_{00}- A_{01}- A_{10})|,
\end{equation}
with $A_{XY}:= P(01|XY) + P(10|XY)$. The deterministic vertices (i.e. the correlations giving deterministic outcomes for all measurements) that saturate inequality (\ref{eq1}) are readily seen to be the following ones \cite{Scarani'06}:
\begin{subequations}
\begin{align}
\mathcal{D}_{1}^{r}= \{P(ab|XY):a(X)=r,b(Y)=r\},\\
\mathcal{D}_{2}^{r}= \{P(ab|XY):a(X)= X\oplus r,b(Y)=r\},\\
\mathcal{D}_{3}^{r}= \{P(ab|XY):a(X)=r,b(Y)= Y\oplus r\},\\
\mathcal{D}_{4}^{r}= \{P(ab|XY):a(X)= X\oplus r,b(Y)= Y\oplus r\oplus 1\},
\end{align}
\end{subequations}
with $r,X,Y\in\{0, 1\}$. Any no-signaling correlation can be expressed as convex mixture of a local correlations and a single extremal nonlocal point on top of each CHSH facet with the representative defined as,
\begin{eqnarray}\label{eq4}
\mathcal{PR}\equiv P(ab|XY):=\begin{cases}
		\frac{1}{2} & \mbox{if}~a\oplus b=XY\\
		0 & \mbox{otherwise}.
		\end{cases}
\end{eqnarray}
This is called PR-correlation (PR-box) as introduced by Popescu and Rohrlich \cite{Popescu'94}. Any no-signaling correlation $P_{NS}\equiv\{P(ab|XY)\}$ can be written as \cite{Scarani'06},
\begin{eqnarray}
P_{NS}=C_1\mathcal{D}_{1}^{0}+C_2\mathcal{D}_{1}^{1}
+C_3\mathcal{D}_{2}^{0}+C_4\mathcal{D}_{2}^{1}
+C_5\mathcal{D}_{3}^{0}\nonumber\\
+C_6\mathcal{D}_{3}^{1}
+C_7\mathcal{D}_{4}^{0}+C_8\mathcal{D}_{4}^{1}
+C_9\mathcal{PR},\label{eq5}
\end{eqnarray}
with $0\le C_i\le 1,~\forall~i$ and  $\sum_{i=1}^9C_{i}=1$. Such a correlation $P_{NS}$ is nonlocal \emph{iff} $P_{NS}\in\mathcal{NS}$ but $P_{NS}\notin\mathcal{L}$.

In the following section we will consider different special forms of NS correlations (\ref{eq5}) and discuss the violation of the Bell-CHSH inequality (\ref{eq1}) in the multi-pair setting of these correlations.
\begin{figure}[t!]
	\centering
	\includegraphics[height=3.3cm,width=7.5cm]{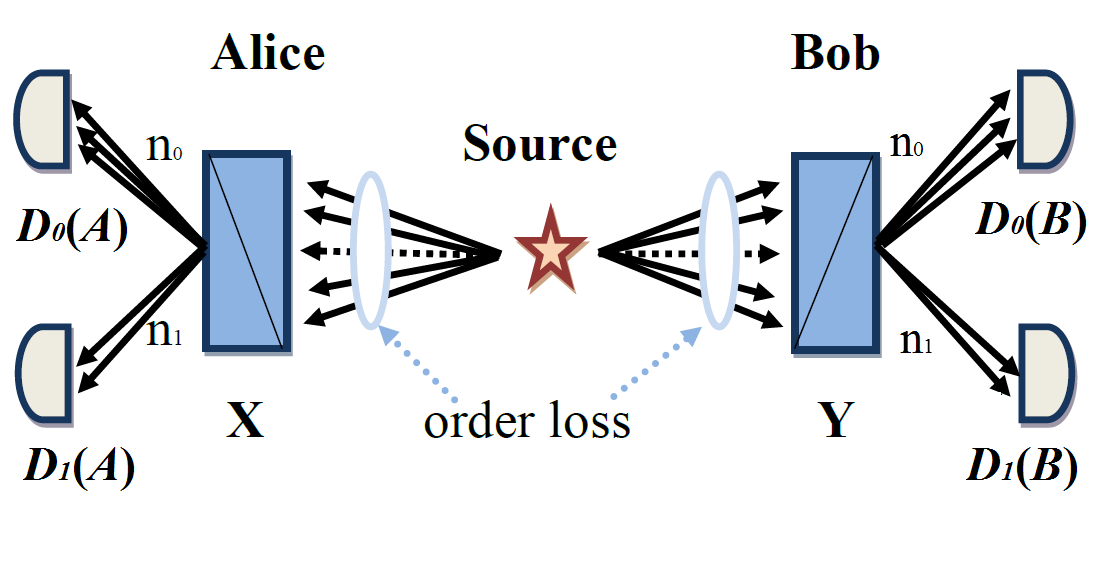}
	\caption{(Color on-line) Multi-pair setup: Source produces $M$ independent pairs of particles. Since information about ordering between Alice's and Bob's particles is lost during their transmission, so they address the beams of particles as a whole. A particle gets in the detector $D_s(\kappa)$ if $s=a\in\{0,1\}$ ($s=b\in\{0,1\}$) is the outcome in the measurement $X\in\{0,1\}$ ($Y\in\{0,1\}$) on $\kappa=A$, i.e Alice's particle ($\kappa=B$, i.e Bob's particle). Few particles ($n_0$ in number) are collected at the detector $D_0(\kappa)$ and the rests at the detector $D_1(\kappa)$, where $\kappa=A,B$.}\label{fig00}
	\vspace{-.3cm}
\end{figure}

\subsection{Multi-pair setting }
Consider that a source produces $M$ independent identical pairs or equivalently $M$ independent sources each producing one and the same pair.  Each of Alice and Bob receives beam of $M$ no of particles (we are assuming that there is no particle loss). Though the pairs are created independently, but in experiment, it is very hard to address them individually (as already discussed earlier). Alice and Bob perform a measurement on the beam of particles they received, i.e. they interact with \emph{all} the particles in same manner as earlier (Alice performs measurement $X\in\{0,1\}$ on \emph{all} the particles she receives and similarly Bob performs $Y\in\{0,1\}$). However, during the interaction the classical information about the identity of the individual pair is lost, i.e., it is not possible to say which particle is correlated with which one (see Fig.\ref{fig00}). Let the correlation of each pair be $P_{NS}=\{P(ab|XY)\}$, and let us denote the global correlation for $M$ pairs as $P_M\equiv P_{NS}^M$. The number of particles collected in two detectors (each on Alice's side and Bob's side) are counted, and let $n_{0}$ and $n_{1}$ are the number of particles counted in two detectors. For perfectly efficient detectors ($\eta = 1$), one has $M=n_{0} + n_{1}$. Our aim is to study the nonlocal strength (particularly, the amount of Bell-CHSH inequality violation) of the global correlation $P_{M}$. For this purpose, Alice and Bob must transform their data, into a binary input-output correlation which we denote in bold letters, i.e., $\{P(\mathbf{ab}|\mathbf{XY})\}$ where $\mathbf{a}, \mathbf{b}, \mathbf{X}, \mathbf{Y}\in\{\mathbf{0}, \mathbf{1}\}$. Here the interactions by Alice and Bob are denoted by bold letters $\mathbf{X}$ and $\mathbf{Y}$ respectively, which imply that they apply the same interaction $X$ or $Y$ on each particle of the incoming beam of particles. And one can get binary outputs by invoking any of the following voting procedures: (a) majority voting, (b) unanimous voting, or (c) any intermediate possibility. According to `majority voting' if the number of particles collected in the detector $D_0(\kappa)$, i.e., $n_0$ is greater than or equal to the number of particles collected in the detector $D_1(\kappa)$ then the outcome will be denoted as $\mathbf{0}$, otherwise the outcome will be denoted as $\mathbf{1}$: 
\begin{eqnarray}
\mbox{Majority voting}\Rightarrow\begin{cases}
		~~n_0\ge n_1~~\longrightarrow~\mathbf{0}, &\\
		\mbox{otherwise}~\longrightarrow~\mathbf{1}. & 
		\end{cases}
\end{eqnarray}
Thus from $M$ independent identical pair of correlations majority voting gives a binary input-output probability distribution $\{P(\mathbf{ab}|\mathbf{XY})\}$. Instead of majority voting one can also follow the voting procedure (b) or (c). However here our aim is to follow such a voting procedure which may exhibit nonlocal behaviour of the given binary input-output correlation even in the macroscopic limit. It may happen that a correlation becomes local in the macroscopic limit under a particular voting protocol, whereas the same correlation exhibits nonlocal behaviour under another voting protocol. It has been shown in \cite{Bancal'08} (see \cite{Brunner'08, Vitelli'10} for experiments that consider majority voting) that the majority voting yields the largest violation and we also checked that the PR-correlation sustains its nonlocal behaviour in the macroscopic limit under majority voting while it becomes local under other voting protocols. For this reason, we consider here the majority voting for our study. In fact, if a NS correlation turns out to be non-local in the macroscopic limit under any one of the aforesaid voting procedures (or, even by using some other counting method in the macroscopic scenario), it will be enough -- according to the notion of macro-realism -- to discard such a correlation as a physical one.

\section{Correlation in multi-pair setting}\label{result}
First we will the consider the $PR$ correlation and then arbitrary no-signaling correlations.
\subsection{PR-correlation}
Before considering the general case of $M$ independent pairs, let us first assume that a source emits two independent pairs of particles each being correlated  according to the PR-correlation of Eq.(\ref{eq4}). Alice (Bob) performs \emph{same} measurement either $X=0$ or $X=1$ ($Y=0$ or $Y=1$) on both the particles she (he) receives. After the measurement they count the number of particles detected on their detectors $D_0(\kappa)$ and $D_1(\kappa)$, $\kappa=A,B$. Then, according to the majority voting condition, they declare their output either as $\mathbf{0}$ or as $\mathbf{1}$, and thus prepare the new binary input-output probability distribution $P(\mathbf{ab}|\mathbf{XY})$. For example, let us consider, both Alice and Bob perform measurement $X=Y=0$ on each particle of their respective beams. The particles can be collected in the detectors in one of the following three ways (see Fig.\ref{fig00} for reference):
\begin{itemize}
\item[(I)] On Alice's side both the particles are detected in the $D_0(A)$ detector. Due to strict correlation of PR-box [see Eq.(\ref{eq4})] both the particles on Bob's side will also be detected in the detector $D_0(B)$. According to the majority vote, both Alice and Bob declare their output as $\mathbf{0}$, i.e. $\mathbf{a}=\mathbf{b}=\mathbf{0}$. The probability of occurrence of this case is $P(\mathbf{a}=\mathbf{0},\mathbf{b}=\mathbf{0}|\mathbf{X}=\mathbf{0},\mathbf{Y}=\mathbf{0})=2!P^2(00|00)/2!$. And for PR-correlation, $ P(00|00)=1/2$.
\item[(II)] On Alice's side both the particles are detected in the  detector $D_1(A)$. Due to similar argument both Alice and Bob declare their output as $\mathbf{1}$, i.e., $\mathbf{a}=\mathbf{b}=\mathbf{1}$, and the probability $P(\mathbf{a}=\mathbf{1},\mathbf{b}=\mathbf{1}|\mathbf{X}=\mathbf{0},\mathbf{Y}=\mathbf{0})$ of occurring this case is $2!P^2(11|00)/2!$, where $P(11|00)=1/2$ for PR correlation.
\item[(III)] On Alice's side one particle is detected in the detector  $D_0(A)$ and the other in the detector $D_1(A)$. Due to strict correlation [see Eq. (\ref{eq4})], same is true on Bob's side. Majority voting condition allows them to declare their output as $\mathbf{0}$. The probability of occurring this case is $P(\mathbf{a}=\mathbf{0},\mathbf{b}=\mathbf{0}|\mathbf{X}=\mathbf{0},\mathbf{Y}=\mathbf{0})=2!P(00|00)P(11|00)/(1!)^2$. 
\end{itemize}
\begin{figure}[t!]
	\centering
	\includegraphics[height=4cm,width=7cm]{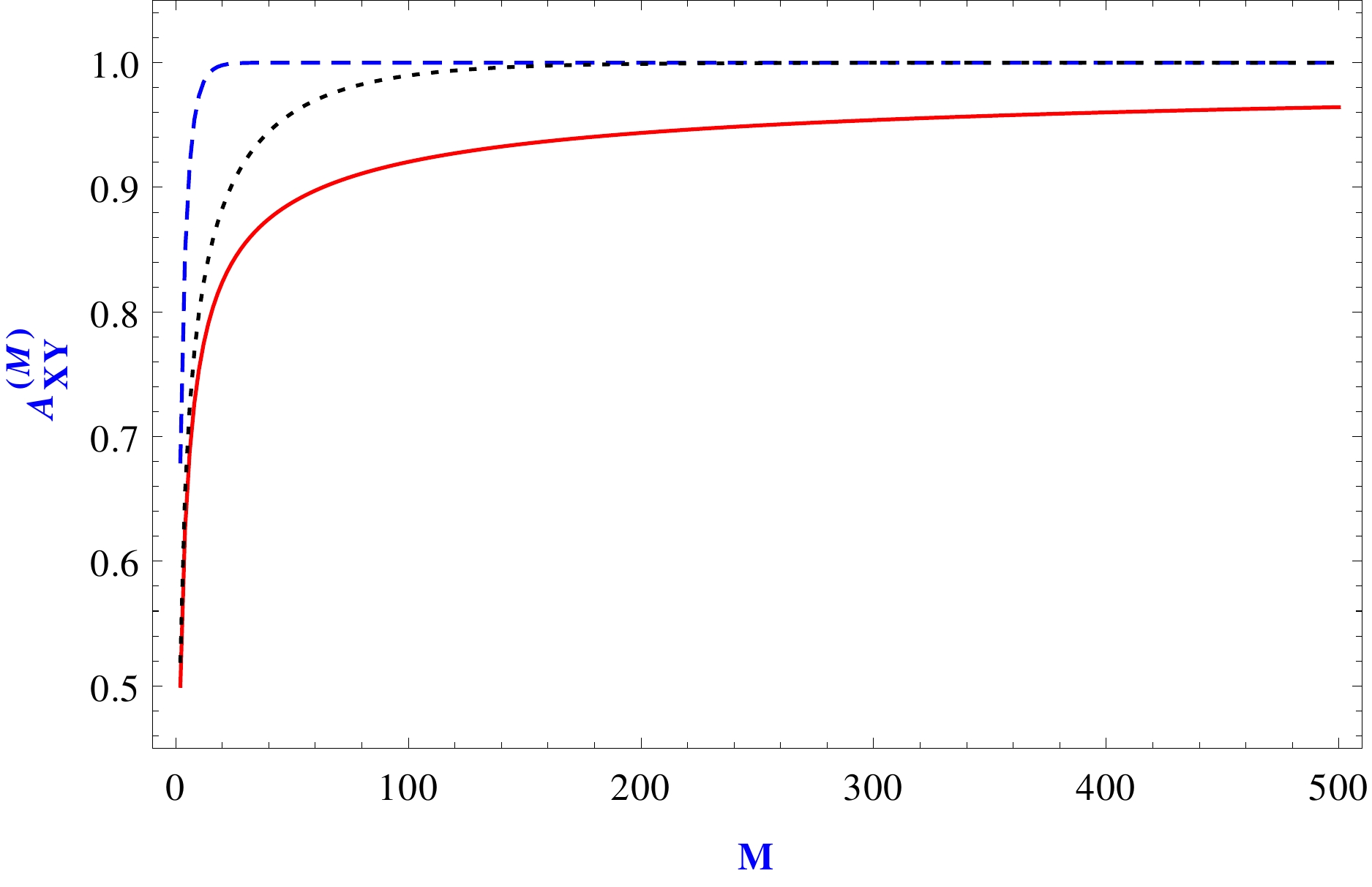}
	\caption{(Color on-line) $A^{(M)}_{\mathbf{XY}}$ for the probability distribution $\mathcal{P}:=\{P(00|XY),P(01|XY),P(10|XY),P(11|XY)\}=(0,\beta,\delta,0)$. Solid curve (red): $\beta=0.5$; dotted curve (black): $\beta=0.4$ dashed curve (blue): $\beta=0.8$.}\label{fig1}
\end{figure}
Thus the new probability distribution for the measurement setting $\mathbf{XY}=\mathbf{00}$ (i.e., $X=0$ on Alice's both particles and $Y=0$ on Bob's both particles) reads,
\begin{eqnarray}
P(\mathbf{00}|\mathbf{00}) &=& 2!\left[\frac{P^2(00|00)}{2!} + \frac{P(00|00)P(11|00)}{(1!)^2}\right],\nonumber\\
P(\mathbf{01}|\mathbf{00})&=&P(\mathbf{10}|\mathbf{00})=0,~P(\mathbf{11}|\mathbf{00})= 2!\left[\frac{P^2(11|00)}{2!}\right].\nonumber
\end{eqnarray}
For the measurement settings $XY=01$, the corresponding \emph{new} probability distribution have the form, 
\begin{eqnarray}
P(\mathbf{00}|\mathbf{01}) &=& 2!\left[\frac{P^2(00|01)}{2!} + \frac{P(00|01)P(11|01)}{(1!)^2}\right],\nonumber\\
P(\mathbf{01}|\mathbf{01})&=&P(\mathbf{10}|\mathbf{01})=0,~P(\mathbf{11}|\mathbf{01})= 2!\left[\frac{P^2(11|01)}{2!}\right],\nonumber
\end{eqnarray}
and similar is the case with $XY=10$. But for $XY=11$ we have,
\begin{eqnarray}
P(\mathbf{00}|\mathbf{11}) = 2!\left[\frac{P(01|11)P(10|11)}{(1!)^2}\right],~~~~P(\mathbf{11}|\mathbf{11}) = 0,~~~~\nonumber\\
P(\mathbf{01}|\mathbf{11})=2!\left[\frac{P^2(01|11)}{2!}\right],~P(\mathbf{10}|\mathbf{11})=2!\left[\frac{P^2(10|11)}{2!}\right].\nonumber 
\end{eqnarray}
To obtain the CHSH value of this new probability distribution we calculate $A^{(2)}_{\mathbf{XY}}=P(\mathbf{01}|\mathbf{XY})+P(\mathbf{10}|\mathbf{XY})$, that in this case become,
\begin{eqnarray}
A^{(2)}_{\mathbf{00}}&=&A^{(2)}_{\mathbf{01}}
=A^{(2)}_{\mathbf{10}}=0,\nonumber\\
A^{(2)}_{\mathbf{11}}&=&2!\left[\frac{P^2(01|11)}{2!}+\frac{P^2(10|11)}{2!}\right].\nonumber
\end{eqnarray}
Here super-index denotes the number of independent pairs used in the experiment. Hence according to Eq.(\ref{eq2}) we have,
\begin{subequations}
\begin{align}\label{8a}
\mathbf{I}^{(2)}_{CHSH}= 2+2\left(A^{(2)}_{\mathbf{11}}-A^{(2)}_{\mathbf{00}}-A^{(2)}_{\mathbf{01}}-A^{(2)}_{\mathbf{10}}\right),\\\label{8b}
=2+2A^{(2)}_{\mathbf{11}}.~~
~~~~~~~~~~~~~~~~~~~~
\end{align}
\end{subequations}
If the source emitting $M$ (let $M$ be even) independent pairs of particle, each paired in PR-correlation, then similar analysis gives, 
\begin{eqnarray}
A^{(M)}_{\mathbf{11}}&=& M!\sum_{j=0}^{\left(\frac{M}{2}-1\right)}\frac{1}{(M-j)!j!}\left[\beta^{(M-j)}\delta^{j} + \beta^{j}\delta^{(M-j)}\right],\nonumber\\
&=&(\beta+\delta)^M-\frac{M!}{(M/2)!}(\beta\delta)^{M/2},\nonumber\\
A^{(M)}_{\mathbf{00}}&=&A^{(M)}_{\mathbf{01}}
=A^{(M)}_{\mathbf{10}}=0,
\label{9}
\end{eqnarray}
where $\beta:= P(01|11)=1/2= P(10|11)=:\delta$. Fig.\ref{fig1} (with $\beta=0.5$) shows that at large $M$ the value of $A^{(M)}_{\mathbf{11}}$ goes close to unity \cite{self} which further implies that $\mathbf{I}^{(M)}_{CHSH}= 2+2A^{(M)}_{\mathbf{11}}\cong 4$ for large $M$, i.e., it reaches the maximum algebraic value of CHSH inequality. Therefore in the macroscopic limit, under majority voting, the PR-correlation does not exhibit classical (more precisely \emph{local}) behavior and hence fails to be considered as a physical correlation.

\subsection{Noisy correlation}\label{3b}
Before going into the explicit examples of NS correlations of Eq.(\ref{eq5}), we first consider different representative cases and study how $A_{\mathbf{XY}}^{(M)}$'s get modified in the macroscopic limit. Here, in all the special cases discussed below, the joint probabilities $P(ab|XY)$ are prescribed for all $XY\in\{00,01,10,11\}$

{\bf Case-1}: For some particular measurement setting $XY\in\{00,01,10,11\}$, let the probability distribution is:
\begin{eqnarray}
P(00|XY)=\alpha,~~~~~P(01|XY)=0,\nonumber\\
P(10|XY)=0,~~~~~P(11|XY)=\gamma,
\end{eqnarray}
with $0\le\alpha,~\gamma\le1,~\alpha+\gamma=1$. From the above discussion (this case is analogous to the the case $XY=00$ of PR box) it is evident that with majority voting, $A^{(M)}_{\mathbf{XY}}=0$ for arbitrary number of pairs $M$.
\begin{figure}[t!]
	\centering
	\includegraphics[height=4cm,width=7cm]{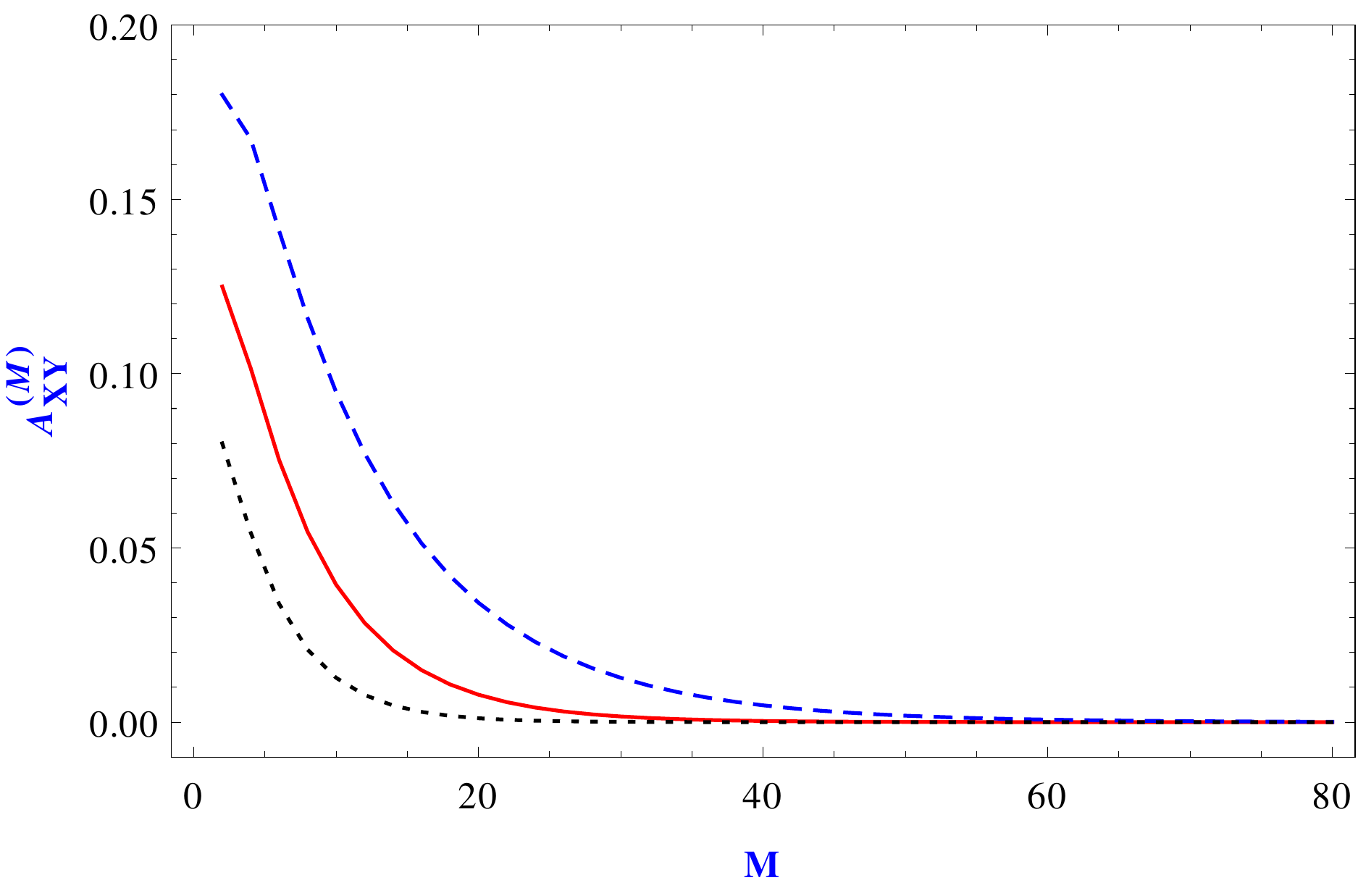}
	\caption{(Color on-line) $A^{(M)}_{\mathbf{XY}}$ for the probability distribution $\mathcal{P}=(\alpha,\beta,\delta,0).$ $\beta=\delta$ and dotted curve (black) $\alpha=0.4$; solid
curve (red):$\alpha=0.5$; dashed curve (blue): $\alpha=0.6$.}\label{fig2}
\end{figure}

{\bf Case-2}: For the measurement setting $XY$ the probability distribution reads,
\begin{eqnarray}
P(00|XY)=0,~~~~~P(01|XY)=\beta,\nonumber\\
P(10|XY)=\delta,~~~~~P(11|XY)=0,
\end{eqnarray}
with $0\le\beta,~\delta\le1,~\beta+\delta=1$. Similar Analysis like PR scenario tells that $A^{(M)}_{\mathbf{XY}}$ look identical to the Eq.(\ref{9}).
For different values of $\beta$, the variation of $A^{(M)}_{\mathbf{XY}}$ with increasing $M$ under majority voting is shown in Fig.\ref{fig1} (with $\beta=0.8$ and $\beta=0.4$), from where it is evident that $A^{(M)}_{\mathbf{XY}}$ approaches to unity at large $M$ limit.

{\bf Case-3}: Let the probability distribution reads:
\begin{eqnarray}
P(00|XY)=\alpha,~~~~~P(01|XY)=\beta,\nonumber\\
P(10|XY)=\delta,~~~~~P(11|XY)=0,
\end{eqnarray}
with $0\le\alpha,~\beta,~\delta\le1,~\alpha+\beta
+\delta=1$. In this case we have,
\begin{eqnarray}
A^{(M)}_{\mathbf{XY}}= M!\sum_{k=0}^{\left(\frac{M}{2}-1\right)}\sum_{j=0}^{\left(\frac{M}{2}-k-1\right)}\frac{\alpha^k}{k!j!(M-k-j)!}\nonumber\\
\times\left[\beta^{(M-k-j)}\delta^{j}
+ \beta^{j}\delta^{(M-k-j)}\right].
\end{eqnarray}
For different choices of $\beta,~\delta$, variations of $A^{(M)}_{\mathbf{XY}}$ with $M$ are plotted in Fig.\ref{fig2}, where it is evident that $A^{(M)}_{\mathbf{XY}}$ approaches to $0$ for large $M$.
\begin{figure}[t!]
	\centering
	\includegraphics[height=4cm,width=7cm]{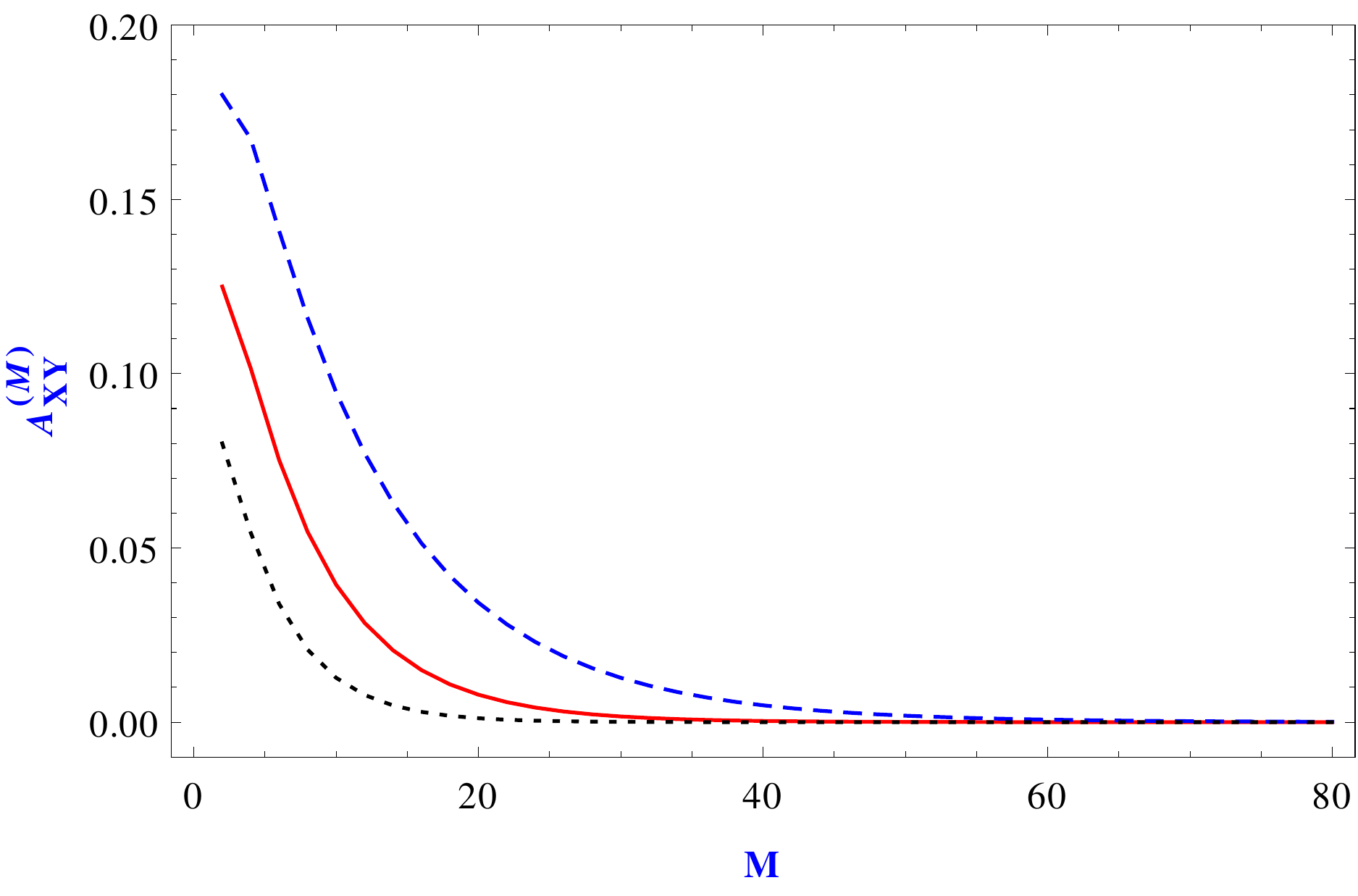}
	\caption{(Color on-line) $A^{(M)}_{\mathbf{XY}}$ for the probability distribution $\mathcal{P}=(0,\beta,\delta,\gamma).$ $\beta=\delta$ and and dotted curve (black): $\gamma=0.6$; solid
curve (red): $\gamma=0.5$; dashed curve (blue): $\gamma=0.4$.}\label{fig3}
\end{figure}

{\bf Case-4}: Here we have,
\begin{eqnarray}
P(00|XY)=0,~~~~~P(01|XY)=\beta,\nonumber\\
P(10|XY)=\delta,~~~~~P(11|XY)=\gamma,
\end{eqnarray}
with $0\le\beta,~\delta,~\gamma\le1,~\beta+\delta
+\gamma=1$. In this case we get,
\begin{eqnarray}
A^{(M)}_{\mathbf{XY}}= M!\sum_{k=0}^{\frac{M}{2}}\sum_{j=0}^{\left(\frac{M}{2}-k-1\right)}\frac{\gamma^{k}}{k!j!(M-k-j)!}\nonumber\\
\times\left[\beta^{(M-k-j)}\delta^{j} + \beta^{j}\delta^{(M-k-j)}\right],
\end{eqnarray}
which is plotted in Fig.\ref{fig3}, which also resembles the same behavior like the Case-3.

\begin{figure}[t!]
	\centering
	\includegraphics[height=4cm,width=7cm]{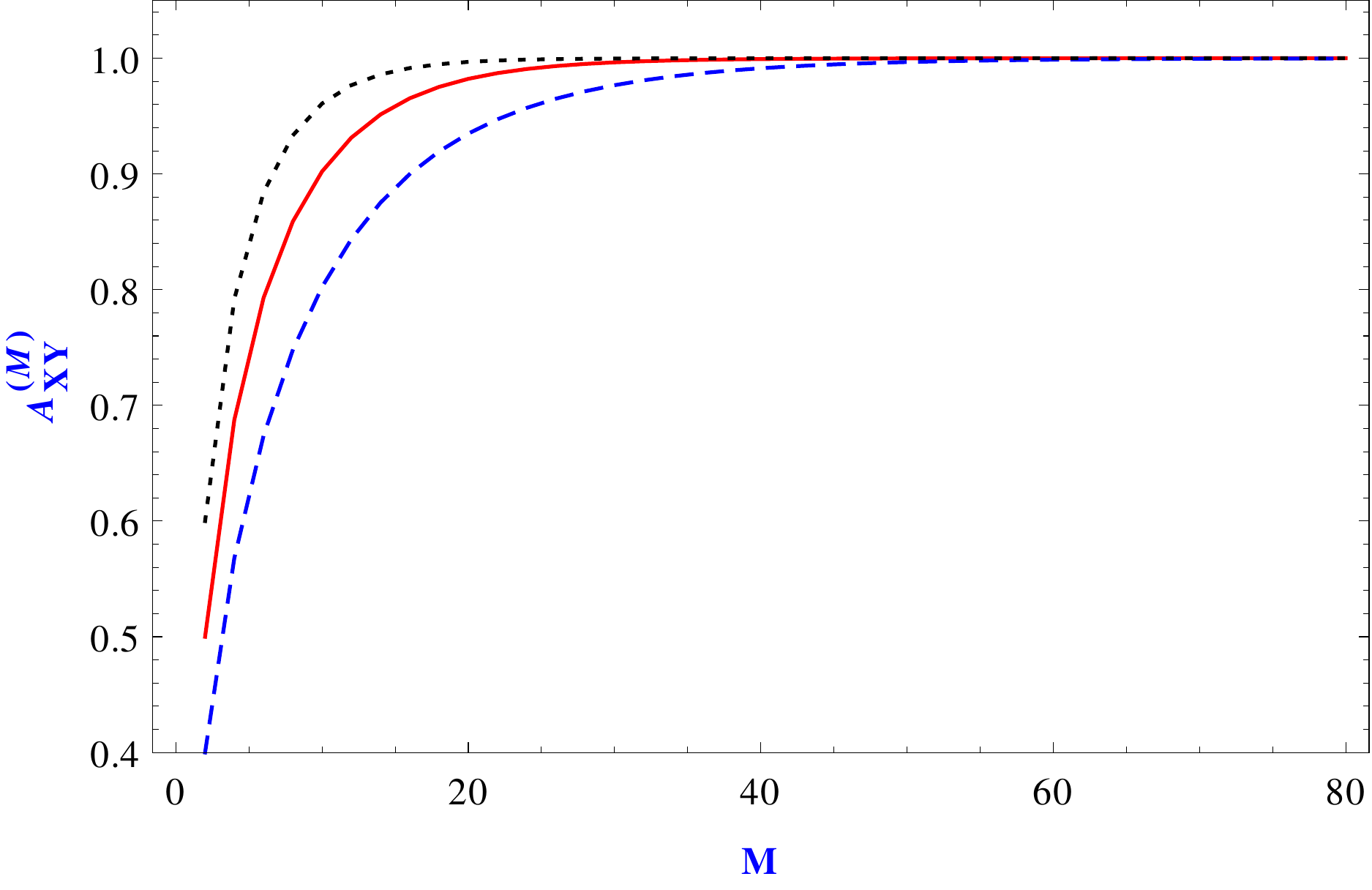}
	\caption{(Color on-line) $A^{(M)}_{\mathbf{XY}}$ for the probability distribution $\mathcal{P}=(\alpha,0,\delta,\gamma).$ $\alpha=\gamma$ and dotted curve (black): $\alpha=0.2$; solid
curve (red): $\alpha=0.25$; dashed curve (blue): $\alpha=0.3$.}\label{fig4}
\end{figure}

{\bf Case-5}: Probability distribution is given by,
\begin{eqnarray}
P(00|XY)=\alpha,~~~~~P(01|XY)=0,\nonumber\\
P(10|XY)=\delta,~~~~~P(11|XY)=\gamma,
\end{eqnarray}
with $0\le\alpha,~\delta,~\gamma\le1,~\alpha+\delta
+\gamma=1$. Here we have,
\begin{eqnarray}
A^{(M)}_{\mathbf{XY}}= M!\sum_{k=0}^{\left(\frac{M}{2}-1\right)}\sum_{j=0}^{\left(\frac{M}{2}-k\right)}\left[\frac{\alpha^{k}\delta^{(M-k-j)}\gamma^{j}}{k!j!(M-k-j)!}\right.\nonumber\\
\left. + \$ \sum_{n=j+1}^{\frac{M}{2}}\frac{\alpha^{k}\delta^{(M-k-n)}\gamma^{n}}{k!n!(M-k-n)!}\right],
\end{eqnarray}
where $\$=1$ when $k+j=\frac{M}{2}$, otherwise $\$=0$. $A^{(M)}_{\mathbf{XY}}$ is plotted in Fig.\ref{fig4},  where $A^{(M)}_{\mathbf{XY}}$ approaches to $1$ for large $M$.

{\bf Case-6}: Probability distribution reads,
\begin{eqnarray}
P(00|XY)=\alpha,~~~~~P(01|XY)=\beta,\nonumber\\
P(10|XY)=0,~~~~~P(11|XY)=\gamma,
\end{eqnarray}
with $0\le\alpha,~\beta,~\gamma\le1,~\alpha+\beta
+\gamma=1$. In this case we get,
\begin{eqnarray}
A^{(M)}_{\mathbf{XY}}= M!\sum_{k=0}^{\left(\frac{M}{2}-1\right)}\sum_{j=0}^{\left(\frac{M}{2}-k\right)}\left[\frac{\alpha^{k}\beta^{(M-k-j)}\gamma^{j}}{k!j!(M-k-j)!}\right.\nonumber\\
\left. + \$ \sum_{n=j+1}^{\frac{M}{2}}\frac{\alpha^{k}\beta^{(M-k-n)}\gamma^{n}}{k!n!(M-k-n)!}\right],
\end{eqnarray}
with $\$=1$ when $k+j=\frac{M}{2}$, otherwise $\$=0$. In this case $A^{(M)}_{\mathbf{XY}}$ looks similar as in Case-5 but $\delta$ is replaced by $\beta$. 

\begin{figure}[t!]
	\centering
	\includegraphics[height=4cm,width=7cm]{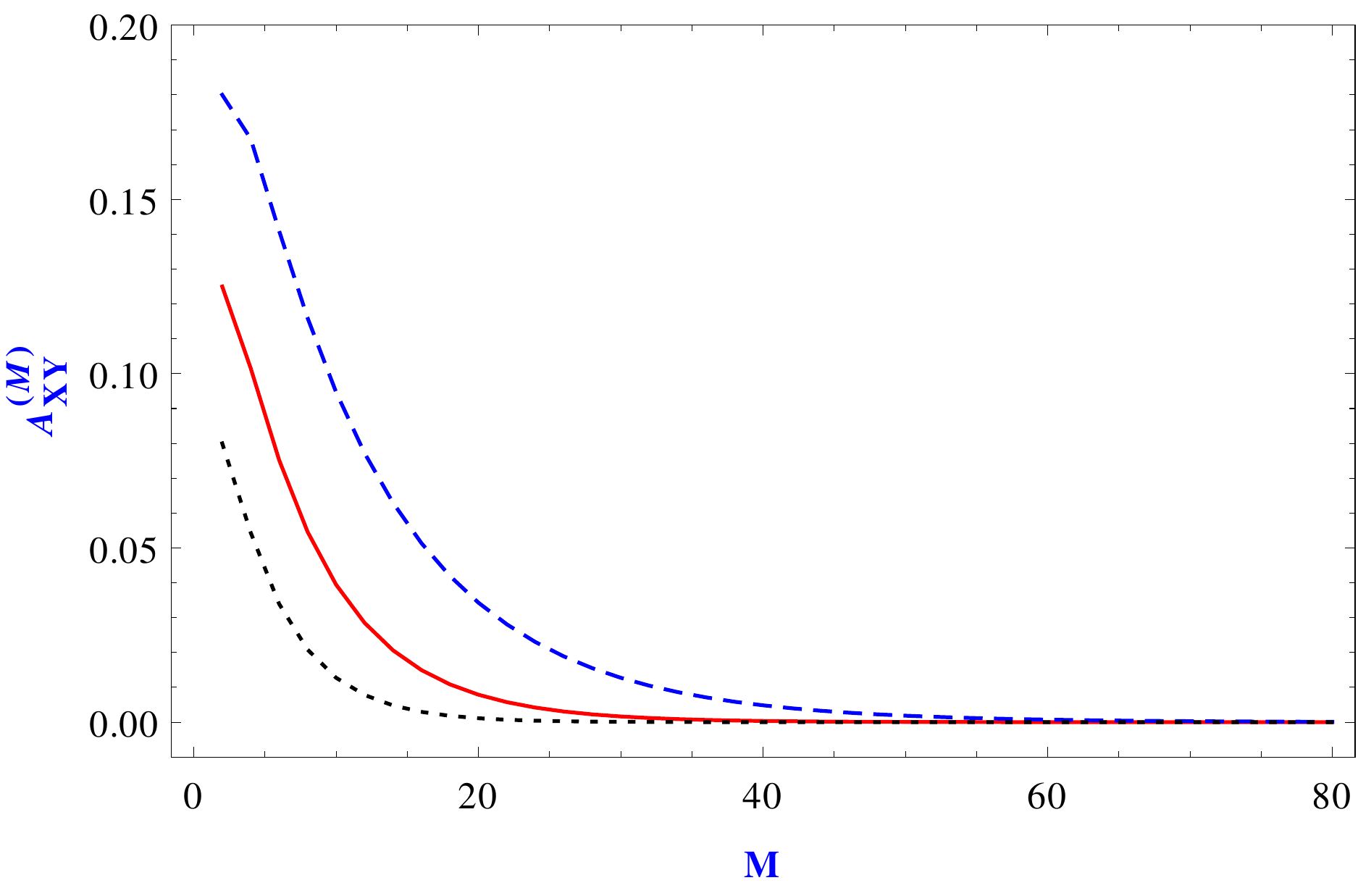}
	\caption{(Color on-line) $A^{(M)}_{\mathbf{XY}}$ for the probability distribution $\mathcal{P}=(\alpha,\beta,\delta,\gamma).$ $\alpha=\gamma,~\beta=\delta$, and dashed curve (blue):
$\alpha=0.2$; solid curve (red): $\alpha=0.25$; dotted curve (black):$\alpha=0.3$.}\label{fig5}
\end{figure}

{\bf Case-7}: Probability distribution is given by:
\begin{eqnarray}
P(00|XY)=\alpha,~~~~~P(01|XY)=\beta,\nonumber\\
P(10|XY)=\delta,~~~~~P(11|XY)=\gamma,
\end{eqnarray}
with $0\le\alpha,~\beta,~\delta,~\gamma\le1,~\alpha
+\beta+\delta+\gamma=1$. In this case we have,
\begin{eqnarray}
A^{(M)}_{\mathbf{XY}}= \sum_{k_1=0}^{\left(\frac{M}{2}-1\right)}\sum_{k_2=0}^{\frac{M}{2}}\sum_{j=0}^{\left(\frac{M}{2}-k_1-k_2-1\right)}\hspace{-.5cm}\frac{M!~\alpha^{k_1}\gamma^{k_2}}{k_1!k_1!j!(M-k_1-k_2-j)!}\nonumber\\
\times\left[\beta^{(M-k_1-k_2-j)}\delta^{j} + \beta^{j}\delta^{(M-k_1-k_2-j)}\right],~~~~~~~~~~~
\end{eqnarray}
which is plotted in Fig.\ref{fig5}, from where it is evident that $A^{(M)}_{\mathbf{XY}}$ approaches to $0$ for large $M$ and consequently $\mathbf{I}_{CHSH}^{(M)}$ becomes $2$. 

We are now in a position to consider some particular nonlocal correlations and thereby test their CHSH values in the macroscopic measurement scenario.

\subsection{Different classes of no-signaling correlations}\label{3c}
In this subsection we will study the nonlocal strengths of different representative classes of no-signaling correlations in macroscopic measurement setting.

{\bf Class-I}: Let the no-signaling probability distribution [see Eq.(\ref{eq5})] be given by
\begin{equation}
P_{NS}= C_9\mathcal{PR} + C_1\mathcal{D}_1^0:=p\mathcal{PR} + (1-p)\mathcal{D}_1^0,
\end{equation}
with $0<C_9(:=p)<1$. The CHSH value of this correlation is $I_{CHSH}= 2 + 2p$. Let the source be emitting $M$ independent pairs of this nonlocal correlation. The joint outcome distributions for the measurement settings $XY=00, 01,10$ are of the form $P(00|XY)=1-p/2,~P(01|XY)=P(10|XY)=0$, and $P(11|XY)=p/2$, which is similar to Case-1 discussed in subsection-\ref{3b}. So, according to majority voting, in the macroscopic measurement scenario $A^{(M)}_{\mathbf{XY}}=0$ for large $M$ limit with $\mathbf{XY}\in\{\mathbf{00},\mathbf{01},\mathbf{10}\}$. For the measurement setting $XY=11$, the probability distribution will be of the form $P(00|11)=1-p,~P(01|11)=P(10|11)=p/2,~P(11|11)=0$ 
---  similar to Case-3 of subsection-\ref{3b}, and hence $A^{(M)}_{\mathbf{11}}=0$ for large $M$. The CHSH value of the microscopic correlation thus becomes: $\mathbf{I}^{(M)}_{CHSH}= 2$. Hence the original microscopic nonlocal correlation becomes local in the macroscopic limit. Same is true for the correlation $P_{NS}= p\mathcal{PR} + (1-p)\mathcal{D}_{1}^{1}$.

{\bf Class-II}: Let the no-signaling probability distribution be of the form
\begin{equation}
P_{NS}= p\mathcal{PR} + (1-p)\mathcal{D}_{2}^{0}.
\end{equation}
Here also the CHSH value is $I_{CHSH}= 2 + 2p$. Outcome
probability distribution for the measurement settings $XY=00, 01$ will be of the form $P(00|XY)=1-p/2,~P(01|XY)=P(10|XY)=0,~P(11|XY)=p/2$, similar to Case-1 of subsection-\ref{3b}, which implies $A^{(M)}_{\mathbf{00}}=A^{(M)}_{\mathbf{01}}=0$ for large $M$. For the measurement setting $10$, the probability distribution is $P(00|01)=P(11|01)=p/2,~P(01|10)=0,~P(10|01)=(1-p)$, which is identical to Case-5 and hence implies $A^{(M)}_{\mathbf{10}}=1$. For the measurement setting $11$, the probability distribution $P(00|11)=0,~P(01|11)=p/2,~P(10|XY)=1-p/2,~P(11|11)=0$ is similar to the Case-2 and hence $A^{(M)}_{\mathbf{11}}=1$. Thus for large $M$, the CHSH value turns out to be,
\begin{equation}
\mathbf{I}^{(M)}_{CHSH}= 2+ 2\left(A^{(M)}_{\mathbf{11}}- A^{(M)}_{\mathbf{10}}- A^{(M)}_{\mathbf{01}}- A^{(M)}_{\mathbf{00}}\right) = 2.
\end{equation}
Similar conclusion holds good for the correlations of the forms $P_{NS}= p\mathcal{PR} + (1-p)\mathcal{D}_{2}^{1}$ and $P_{NS}= p\mathcal{PR}+(1-p)\mathcal{D}_{s}^{r}$ for $s=3,4,$ and $r=0,1$.

{\bf Class-III}: Let the probability distribution be given by,
\begin{equation}
P_NS= p_1\mathcal{PR} + p_2\mathcal{D}_{1}^{0} + p_3\mathcal{D}_{1}^{1}, 
\end{equation}
with $0<p_i<1$, $\sum p_i=1$ and the CHSH value is $I_{CHSH}= 2 + 2p_1$. The outcome probability distribution for the measurement settings $XY=00, 01,10$ is of the form of Case-1 and for $XY=11$, it is of the form of Case-7 of subsection-\ref{3b}, implying $A^{(M)}_{\mathbf{XY}}=0$ for all $\mathbf{XY}$. This further implies that $\mathbf{I}_{CHSH}^{(M)}=2$ for large $M$. 

{\bf Class-IV}: Let the probability distribution be of the form,
\begin{equation}
P_{NS}= p_1\mathcal{PR} + p_2\mathcal{D}_{2}^{0} + p_3\mathcal{D}_{2}^{1},
\end{equation}
with $0<p_i<1,~\sum p_i=1$, and the CHSH value is therefore: $I_{CHSH}= 2 + 2p_1$. For the measurement settings $XY=00,01$ the outcome distribution will be of the form similar to Case-I while for $XY=10$ it resembles Case-7 of subsection-\ref{3b}, and thus it implies $A^{(M)}_{\mathbf{XY}}=0$ for measurement settings $\mathbf{XY}\in\{\mathbf{00},\mathbf{01},\mathbf{10}\}$, for large $M$. On the other hand, for the measurement setting $XY=11$, outcome distribution will be of the form of Case-2 of subsection-\ref{3b}, implying $A^{(M)}_{\mathbf{11}}=1$. This further gives that at large $M$ we have $\mathbf{I}^{(M)}_{CHSH}=4$. Similar conclusion holds good for the correlations belonging to the classes $P_{NS}= p_1\mathcal{PR} + p_2\mathcal{D}_{s}^{0} + p_3\mathcal{D}_{s}^{1}$ with $s=3,4$. Therefore, for these classes of correlations, the original \emph{weak} microscopic nonlocality become maximally nonlocal in the macroscopic limit under majority voting condition. 

{\bf Class-V}: Let the probability distribution be given by,
\begin{equation}
P_{NS}= p_1\mathcal{PR} + p_2\mathcal{D}_{1}^{0} + p_3\mathcal{D}_{2}^{0}+p_4\mathcal{D}_{3}^{0} + p_5\mathcal{D}_{4}^{0},
\end{equation}
with $0<p_i<1,~\sum p_i=1$ and $I_{CHSH}= 2 + 2p_1$.The outcome probability distribution for the measurement settings $XY=00, 01$ is similar to Case-6, while for the measurement setting $XY=10$ and $XY=11$, they are similar to the Case-5 and Case-3 of subsection-\ref{3b}, respectively. So, for large $M$ we have,
\begin{equation}
\mathbf{I}^{(M)}_{CHSH}=2+2\left(A^{(M)}_{\mathbf{XY}}-A^{(M)}_{\mathbf{XY}}-A^{(M)}_{\mathbf{XY}}-A^{(M)}_{\mathbf{XY}}\right)=-4.
\end{equation}
Thus, in this case also the original \emph{weak} microscopic nonlocal correlations become maximally nonlocal (i.e. CHSH value $4$) in the macroscopic limit according to majority voting condition. Similar result holds for the other correlations of the forms
$P_{NS}= p_1\mathcal{PR} + p_2\mathcal{D}_{1}^{r} + p_3\mathcal{D}_{2}^{t}+p_4\mathcal{D}_{3}^{u} + p_5\mathcal{D}_{4}^{v}$, with $r,t,u,v\in\{0.1\}$.

In the line of the aforementioned analysis one can consider \emph{any} of $2-2-2$ NS correlation of Eq.(\ref{eq5}) and can find its nonlocal strength in the macroscopic limit. 

\section{Unphysical correlations: nonlocality distillation, Information Causality}\label{comparison}
If one \emph{believes} that Nature does not allow to perform all distributed computations with a trivial amount of communication, or one \emph{believes} in the principle that the amount of information that an observer ( say, Bob) can gain about a data set belonging to another observer (say, Alice), using all of his local resources (which may be correlated with her resources) and using classical communication obtained from Alice, is bounded by the information volume of the communication, then, under the aforesaid belief, not all no-signaling correlations can be considered as physical. In this context, \emph{nonlocality distillation} and \emph{information causality} principle are two well known tests to determine whether a given no-signaling correlation is unphysical.

\emph{Nonlocality distillation}: This idea has been proposed by Forster \emph{et al} \cite{Forster'09}. Starting  from  several  copies  of  a nonlocal  box  with  a  given  CHSH  value (greater than $2$), it is possible via wiring (classical circuitry to produce a new binary-input/binary-output box, or in other way, to say post-processing of the data but without any communication) to obtain a final box which has a larger CHSH value. Using this idea, in Ref.\citep{Brunner'09}, the authors have identified a specific class of post-quantum nonlocal boxes that make communication complexity trivial, and therefore such correlations are very unlikely to exist in Nature. In our analysis, we find that correlations belonging to Class-I and Class-II of subsection-\ref{3c} are local in the macroscopic measurement scenario under majority voting. However, as shown in \cite{Allcock'09}, these correlations can be distilled arbitrarily close to the maximally nonlocal correlation, implying \emph{trivial communication complexity}; and hence such correlations are considered to be unphysical (according the aforesaid belief).
\begin{figure}[t!]
	\centering
	\includegraphics[height=7cm,width=7cm]{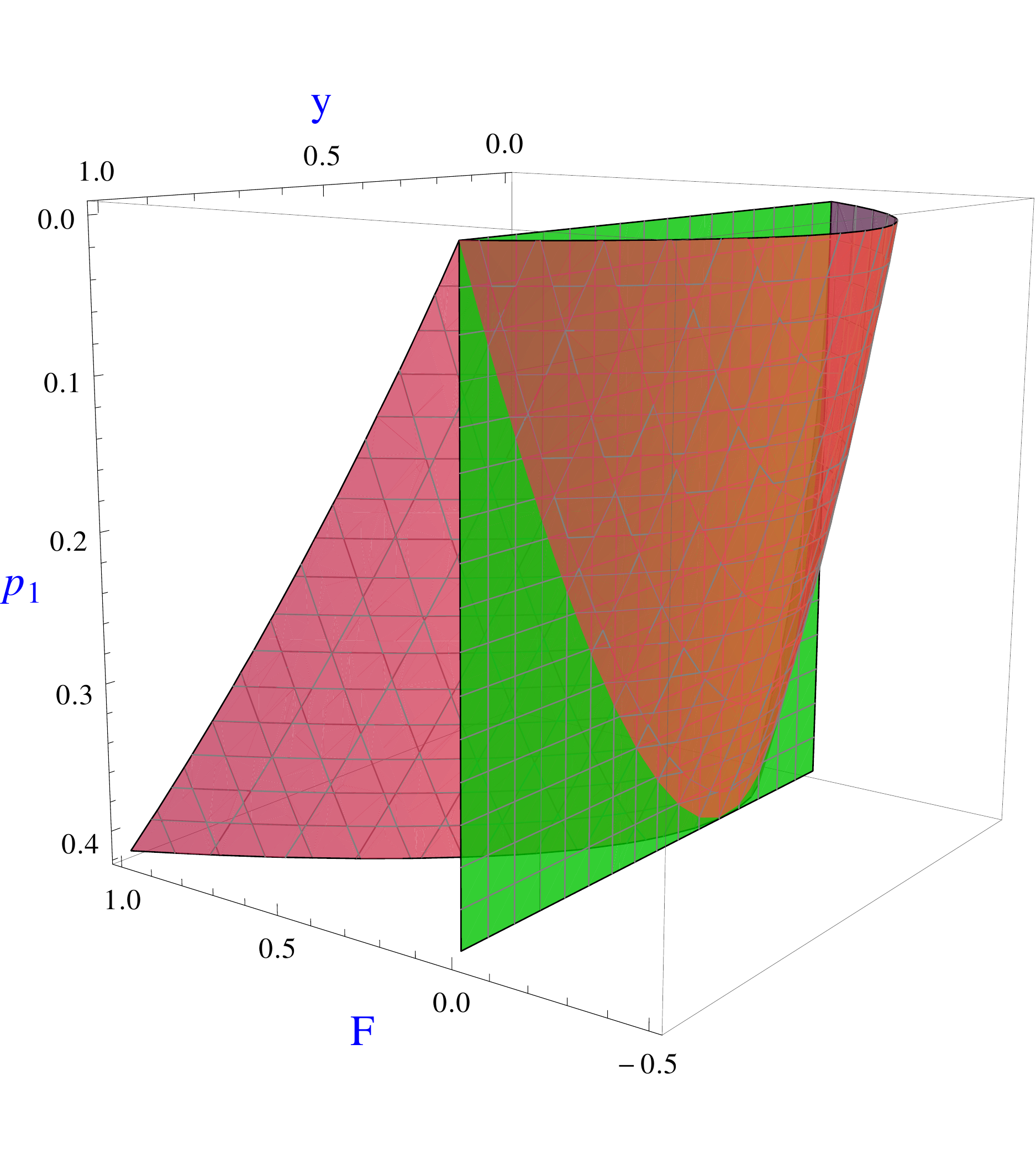}
	\caption{(Color on-line) Red surface represents the function $F(p_1,y)$ of Eq.(\ref{eq32}). The green surface represents $F=0$ surface. The points $(p_1,y)$ at the right side of the green surface satify the necessary condition of IC.}\label{fig6}
\end{figure}

\emph{Information Causality}: Pawlowski \emph{et al} have proposed the principle of information causality (IC) as a generalization of no-signaling principle. It can be formulated quantitatively through an information processing game played between two parties \cite{Pawlowski'09}.  If Alice communicates $m$ bits to Bob, the total information obtainable by Bob, using all his local resources (which may be correlated with Alice's resources) and the classical communications from her, cannot be greater than $m$. For $m = 0$, IC reduces to the standard no-signaling principle. Both classical and quantum correlations have been proved to satisfy the IC principle. Furthermore it has been shown that, if Alice and Bob share arbitrary two-input and two-output no-signaling correlations, then by applying a protocol by Van Dam \cite{vanDam'05} and Wolf \emph{et al} \cite{Wolf'05}, one can derive a necessary condition for respecting the IC principle which can be expressed as,
\begin{equation}
E_{1}^2 + E_{2}^2 \le 1.
\end{equation}
where $E_i = 2Q_i - 1$ for $i=1,2$ and $Q_1 = \frac{1}{2}[P(a=b|00)+ P(a=b|10)]$, $Q_2 = \frac{1}{2}[P(a=b|01)+ P(a\neq b|11)]$. 

For the probability distributions belonging to the Class-V of subsection-\ref{3c}, we have $E_1= 1-(p_3+p_5)$ and $E_2= 1-(p_2+p_4)$. The necessary condition of IC thus implies,
\begin{equation}
p_1^2 -2(p_3+p_5)(p_2+p_4) \le 0,
\end{equation}
i.e., the probability distributions belonging to the Class-V of subsection-\ref{3c} will satisfy the necessary condition of IC as long as the function $F:=p_1^2 -2(p_3+p_5)(p_2+p_4) $ is not positive. Since the Bell-CHSH expression for the probability distributions belonging to the Class-V is $2+2p_1$, they violate the Tsirelson’s bound if $p_1>\sqrt{2}-1$ and hence are not quantum. So we are  interested in the range $0\le p_1\le \sqrt{2}-1$. Now letting $y=p_3+p_5$ (clearly $0\le y\le 1$) and using the probability normalization condition, i.e., $p_1+p_2+p_3+p_4+p_5=1$, we get  
\begin{equation}
F=p_1^2-2y+2p_1y+2y^2.\label{eq32}
\end{equation}
We plot the function $F(p_1,y)$ in Fig.\ref{fig6} which shows that, in our interested ranges of parameter $p_1$ ({\it i.e } $0\le p_1\le \sqrt{2}-1$), there exist correlations which satisfy the necessary condition of IC. Therefore the necessary condition of IC fails to identify those correlations as unphysical. However, our earlier analysis points them out as unphysical one since these correlations show extreme non local behaviour (i.e Bell-CHSH value $4$) and hence fail to exhibit the expected classical feature (i.e. the local behaviour of the correlation) in the macroscopic limit even though, at single-copy level they do not violate the Tirelson's bound. 

\section{Concluding remarks}\label{conclusion}
Identifying the set of all quantum correlations is a very  important problem in the research area of quantum foundation. This also has practical relevance since nonlocal correlations are resources for various device-independent tasks. In the last few years, different approaches, based on information theoretic or physical principles, have been proposed to identify the quantum correlations \cite{Pawlowski'09,Navascues'09}. Whereas in \cite{Pawlowski'09}, the authors introduced an information theoretic principle, namely IC, in \cite{Navascues'09} the authors introduced a physical principle, namely ML. In this paper we take a different approach which is closer to the second one. Whereas according to ML, the coarse-grained extensive observations of macroscopic sources of $M$ independent particle pairs should admit a local hidden variable model in the limit $M\rightarrow\infty$, we have considered the majority voting approach (like \cite{Bancal'08}) to get a new probability distribution from $M$ independent particle pairs and demand that in the limit $M\rightarrow\infty$ this new correlation should behave locally. For the simplest scenario ($2-2-2 $ case) we show how one can characterize which correlations become local and which are not. Correlations exhibiting nonlocal behavior in large $M$ limit are sure to be unphysical. We also find that for some set of correlations, our method is better than the necessary criterion of the IC principle in identifying them as unphysical ones. Moreover our approach identifies some no-signaling correlations each of which does satisfy the Tsirelson’s bound \emph{non-maximally}, but gives rise to maximum non-locality in the macroscopic measurement setup. As a future work it will be interesting to extend this study for more general input-output correlations rather than just $2-2-2$ scenario.

{\bf Acknowledgments}: The authors would like to thank Guruprasad Kar for many stimulating discussions. SK would like to acknowledge the visit at The Institute of Mathematical Sciences, Chennai, where part of this work has been done. SK thanks UGC for financial support through Minor Research Project [Grant No. PSW-177/14-15(ERO)].



\begin{thebibliography}{99}
\bibitem{self} For large $M$, using Sterling formula, we have $(\beta+\delta)^M-\frac{M!}{(M/2)!}(\beta\delta)^{(M/2)}\cong 1-M\ln M+M\ln(M/2)-M\ln((\beta\delta)^{1/2})$.
\end{thebibliography}
\end{document}